\documentclass[iop,natbib]{emulateapj}

\usepackage[]{natbib}

\newcommand{\pmra}{$\mu_{\alpha}$}
\newcommand{\pmdec}{$\mu_{\delta}$}

\newcommand{\name}{W0720}
\newcommand{\nameA}{W0720A}
\newcommand{\nameB}{W0720B}
\newcommand{\hip}{\emph{Hipparcos}}

\slugcomment{Astrophysical Journal Letters, 800, L17}
\shorttitle{Oort Cloud Perturber}
\shortauthors{Mamajek et al.}
\begin{document}

\title{The Closest Known Flyby of a Star to the Solar System}

\author{
Eric E. Mamajek\altaffilmark{1},
Scott A. Barenfeld\altaffilmark{2},
Valentin D. Ivanov\altaffilmark{3,4},
Alexei Y. Kniazev\altaffilmark{5,6,7},\\
Petri V\"ais\"anen\altaffilmark{5,6},
Yuri Beletsky\altaffilmark{8},
Henri M. J. Boffin\altaffilmark{3}}
\altaffiltext{1}{Department of Physics \& Astronomy, University of Rochester, Rochester, NY 14627, USA}
\altaffiltext{2}{Department of Astronomy, California Institute of Technology, Pasadena, CA 91125, USA}
\altaffiltext{3}{European Southern Observatory, Av. Alonso de Cordova 3107, 19001 Casilla, Santiago 19, Chile}
\altaffiltext{4}{European Southern Observatory, Karl-Schwarzschild-Str. 2, 85748 Garching bei M\"unchen, Germany}
\altaffiltext{5}{South African Astronomical Observatory, PO Box 9, 7935 Observatory, Cape Town, South Africa}
\altaffiltext{6}{Southern African Large Telescope Foundation, PO Box 9, 7935 Observatory, Cape Town, South Africa}
\altaffiltext{7}{Sternberg Astronomical Institute, Lomonosov Moscow State University, Moscow, Russia}
\altaffiltext{8}{Las Campanas Observatory, Carnegie Institution of Washington, Colina el Pino, Casilla 601 La Serena, Chile}
\email{emamajek@pas.rochester.edu}

\begin{abstract}
Passing stars can perturb the Oort Cloud, triggering comet showers and
potentially extinction events on Earth.
We combine velocity measurements for the recently discovered, nearby,
low-mass binary system WISE J072003.20-084651.2 (``Scholz's star'') to
calculate its past trajectory.
Integrating the Galactic orbits of this $\sim$0.15 M$_{\odot}$ binary
system and the Sun, we find that the binary passed within only
52$^{+23}_{-14}$ kAU (0.25$^{+0.11}_{-0.07}$ parsec) of the Sun
70$^{+15}_{-10}$ kya (1$\sigma$ uncertainties), i.e. within the outer
Oort Cloud.
This is the closest known encounter of a star to our solar system with
a well-constrained distance and velocity.
Previous work suggests that flybys within 0.25 pc occur infrequently
($\sim$0.1 Myr$^{-1}$).
We show that given the low mass and high velocity of the binary
system, the encounter was dynamically weak.
Using the best available astrometry, our simulations suggest that the
probability that the star penetrated the outer Oort Cloud is
$\sim$98\%, but the probability of penetrating the dynamically active
inner Oort Cloud ($<$20 kAU) is $\sim$10$^{-4}$.
While the flyby of this system likely caused negligible impact on the
flux of long-period comets, the recent discovery of this binary
highlights that dynamically important Oort Cloud perturbers may be
lurking among nearby stars.
\end{abstract}


\section{Introduction\label{sec:intro}}

Perturbations by passing stars on Oort cloud comets have previously
been proposed as the source of long-period comets visiting the
planetary region of the solar system \citep{Oort50, Biermann83,
  Weissman96, Rickman14}, and possibly for generating Earth-crossing
comets that produce biological extinction events \citep{Davis84}.
Approximately 30\%\, of craters with diameters $>$10 km on the Earth
and Moon are likely due to long-period comets from the Oort Cloud
\citep{Weissman96}.  Periodic increases in the flux of Oort cloud
comets due to a hypothetical substellar companion have been proposed
\citep{Whitmire84}, however recent time series analysis of terrestrial
impact craters are inconsistent with periodic variations
\citep{Bailer-Jones11}, and sensitive infrared sky surveys have
yielded no evidence for any wide-separation substellar companion
\citep{Luhman14}.  A survey of nearby field stars with \hip\,
astrometric data \citep{Perryman97} by \citet{Garcia-Sanchez99}
identified only a single candidate with a pass of within 0.9 pc of the
Sun (Gl 710; 1.4 Myr in the future at $\sim$0.34 pc), however it is
predicted that $\sim$12 stars pass within 1 pc of the Sun every Myr
\citep{Garcia-Sanchez01}. A recent analysis by \citet{Bailer-Jones14}
of the orbits of $\sim$50,000 stars using the revised \hip\,
astrometry from \citet{vanLeeuwen07}, identified four \hip\, stars
whose future flybys may bring them within 0.5 pc of the Sun (however
the closest candidate HIP 85605 has large astrometric uncertainties;
see discussion in \S3).\\

A low-mass star in the solar vicinity in Monoceros, WISE
J072003.20-084651.2 (hereafter \name\, or ``Scholz's star'') was
recently discovered with a photometric distance of $\sim$7 pc and
initial spectral classification of M9\,$\pm$\,1 \citep{Scholz14}. This
nearby star likely remained undiscovered for so long due to its
combination of proximity to the Galactic plane ($b$ = +2$^{\circ}$.3),
optical dimness ($V$ = 18.3), and low proper motion ($\sim$0''.1
yr$^{-1}$).  The combination of proximity and low tangential velocity
for \name\, ($V_{tan}$ $\simeq$ 3 km\,s$^{-1}$) initially drew our
attention to this system. If most of the star's motion was radial, it
was possible that the star may have a past or future close pass to the
Sun. Indeed, \citet{Burgasser14} and \citet{Ivanov14} have recently
reported a high positive radial velocity. \citet{Burgasser14} resolved
\name\, as a M9.5+T5 binary and provided a trigonometric parllax
distance of 6.0$^{+1.2}_{-0.9}$ pc. Here we investigate the trajectory
of the \name\, system with respect to the solar system, and
demonstrate that the star recently ($\sim$70,000 years ago) passed
through the Oort Cloud.

\section{Data \& Analysis\label{sec:data}}

We obtained medium resolution spectroscopy of \name\, on UT 17 and 19
November 2013 with the Southern African Large Telescope (SALT) and
Magellan telescopes, in the optical and near-infrared respectively.
As summarized in \citet{Ivanov14}, we estimated the spectral type of
\name\, to be L0\,$\pm$\,1, and measured a radial velocity of
77.6\,$\pm$\,2.5 km\,s$^{-1}$. Recent adaptive optics imaging and high
resolution spectroscopy by \citet{Burgasser14} indicated that the star
is actually a low-mass binary classified as M9.5+T5, with projected
separation 0.8 AU, and a multi-epoch mean radial velocity of
83.1\,$\pm$\,0.4 km\,s$^{-1}$.  Given the radial velocity accuracy,
number of reported epochs (11), and improved astrometric solution
provided by the recent \citet{Burgasser14} study, we simply adopt
their parameters for our calculations: $\alpha$ = 07:20:03.21 (ICRS;
epoch 2014.0), $\delta$ = -08$^{\circ}$46\arcmin 51\arcsec .83 (ICRS; epoch
2014.0), $\mu_{\alpha}$ = -40.3\,$\pm$\,0.2 mas\,yr$^{-1}$,
$\mu_{\delta}$ = -114.8\,$\pm$\,0.4 mas\,yr$^{-1}$, $\varpi$ =
166\,$\pm$\,28 mas, $d$ = 6.0$^{+1.2}_{-0.9}$ pc, $v_{radial}$ =
83.1\,$\pm$\,0.4 km\,s$^{-1}$, $v_{total}$ = 83.2\,$\pm$\,0.4
km\,s$^{-1}$ (all 1$\sigma$ uncertainties).

Using the Galactic velocity ellipsoids and population normalizations
from \citet{Bensby03}, we estimate that \name\, has probabilities of
77.3\%, 22.5\%, and 0.2\%\, of belonging to the thin disk, thick disk,
and halo Galactic populations, respectively.  The spectral
classification indices are consistent with a thin disk star of
approximately solar composition \citep{Burgasser14} \citep[classified
  ``dM'' based on $\zeta$ = 1.034\,$\pm$\,0.018 ;][]{Lepine07}.
Classification as a metal-poor subdwarf is ruled out, and given the
correlation between metallicity and kinematic properties for M-type
stars \citep{Savcheva14}, it is unlikely that \name\, belongs to
either the thick disk or halo.  Among 890 nearby solar-type stars from
a chromospheric activity-velocity catalog \citep{Jenkins11}, only 3
have Galactic velocity components within $\pm$10 km\,s$^{-1}$ of
\name's velocity (HIP 51500, 63851, 117499), and all three are less
chromospherically active than our Sun (log\,R'$_{HK}$ $\leq$ -4.95)
and have inferred chromospheric ages in the range 4-8 Gyr
\citep{Mamajek08}.  \name's velocity is similar to that of the
Hercules dynamical stream, a kinematic group of stars of heterogenous
ages and composition, likely perturbed to the solar circle from
smaller Galactocentric radii due to dynamical interactions with the
Galactic bar \citep{Bensby07}.  Based on its status as an old thin
disk star, consideration of the ages of Sun-like stars of similar
velocity, and the isochronal ages for other Hercules stream members
\citep{Bensby14}, we adopt an age for \name\, of 3-10 Gyr (2$\sigma$
range).  Interpolating the mass-age estimates for the \name\,
components from \cite{Burgasser14} for solar composition, this age
range maps to component masses of M$_A$ = 86\,$\pm$\,2 M$_{Jup}$ and
M$_B$ = 65\,$\pm$\,12 M$_{Jup}$ (2$\sigma$ uncertainties).  The
hydrogen-burning mass limit for stars is near $\sim$75 M$_{Jup}$
\citep{Saumon08, Dieterich14}, hence \nameA\, is probably a low-mass
star, and \nameB\, is probably a brown dwarf.\\

We integrated the orbit of \name\, and the Sun with a realistic
Galactic gravitational potential using the NEMO Stellar Dynamics
Toolbox \citep[][]{Teuben95,Barenfeld13}.  With the current velocity
data, we simulated 10$^4$ orbits of \name\, and the Sun, sampling the
observed astrometric values and observational uncertainties for
\name\, using Gaussian deviates.  From these 10$^4$ simulations which
take into account the observational uncertainties, we find that
\name\, passed as close as $\Delta$ = 0.252$^{+0.111}_{-0.068}$ pc
(1$\sigma$; $^{+0.317}_{-0.110}$ pc 2$\sigma$) or
52.0$^{+22.9}_{-13.9}$ kAU (1$\sigma$; $^{+65.5}_{-22.7}$ kAU
2$\sigma$) of the Sun.  The time of closest approach was
70.4$^{+14.7}_{-9.8}$ (1$\sigma$; $^{+36.2}_{-17.6}$ 2$\sigma$)
kya\footnote{kya = thousand years ago.}. Figure 1 shows the
distributions of the closest approach separations and times for
\name\, and the Sun for the 10$^4$ orbit simulations. The median
nearest pass position from the simulations was ($X, Y, Z$) = (-0.122,
+0.120, +0.185) pc (here quoted in a comoving frame centered on the
solar system barycenter, where $X$ is in the current direction of the
Galactic center, $Y$ is in the current direction of Galactic rotation,
and $Z$ is towards the current north Galactic pole), with approximate
1$\sigma$ uncertainties of ($\sigma_X$, $\sigma_Y$, $\sigma_Z$) =
(0.047, 0.045, 0.068) pc, corresponding to celestial position
($\alpha_{ICRS}$, $\delta_{ICRS}$ $\simeq$
170$^{\circ}$\,$\pm$\,29$^{\circ}$,
+68$^{\circ}$\,$\pm$\,14$^{\circ}$; 1$\sigma$ unc.), in the vicinity
of Ursa Major.  As a check, we also calculated a linear trajectory
that ignores the Galactic potential, finding that \name's closest
approach was 70.7 kya at 0.25 parsec at a barycentric position in
Galactic coordinates of ($X, Y, Z$ = -0.125, +0.119, +0.185 pc).  Note
that given the short timespan, the predictions of the time and
position of the \name-Sun minimum pass from the linear trajectory
agrees with the more accurate orbit integration to better than 2.5\%.
At its closest pass \name\, would have had proper motion exceeding any
known star: 70'' yr$^{-1}$ (i.e. capable of traversing a full moon in
26 years; cf. the current highest proper motion star, Barnard's star,
with 10.3\arcsec\,yr$^{-1}$).\\

The predicted Galactic coordinates of the nearest pass of the binary
system ($\ell$, $b$ = 135$^{\circ}$\,$\pm$\,15$^{\circ}$,
47$^{\circ}$\,$\pm$\,13$^{\circ}$) is near one of the two strong peaks
in the longitude of aphelia distribution of new class I comets from
the Oort Cloud ($\ell$ = 135$^{\circ}$\,$\pm$\,15$^{\circ}$)
\citep{Matese99}.  However this appears to be coincidence, as any
comets on eccentricity $\sim$ 1 orbits from the vicinity of the
nearest pass $\sim$70 kya would have periods of $\sim$4 Myr, and hence
require $\sim$2 Myr to reach the inner solar system.  Also, the two
primary peaks in the distribution of longitudes of aphelia for
long-period comets are reasonably explained by the effects of the
solar motion and Galactic tide \citep{Feng14}.  Among 10$^4$
simulations, the nearest past separation between \name\, and the Sun
was $\Delta$ = 0.087 pc (18.0 kAU), and this was the only simulation
that brought \name\, within the classical boundary of the dynamically
active inner Oort Cloud (``Hills Cloud'') \citep{Hills81, Weissman96}
at $a$ $<$ 20 kAU.  Approximately 79\%\, of the simulations brought
\name\, within 0.337 pc of the Sun, the previously closest estimated
pass of a known star to the solar system \citep[Gliese 710; 0.337 pc,
  1.4 Myr in the future;][]{Garcia-Sanchez01}.  Approximately
99.96\%\, of the simulated trajectories brought the star within the
Sun's tidal radius of 1.35 pc \citep{Mamajek13}, and 98\%\, of the
simulated trajectories brought the star within the maximal range of
semi-major axes for retrograde orbiting Oort Cloud comets ($\sim$120
kAU) \citep{Garcia-Sanchez99}.  The rarity of stellar passes with such
a small impact parameter can be assessed from the analysis of
\citet{Garcia-Sanchez01}.  Extrapolating the power-law distribution of
minimum separations versus cumulative number of encounters per Myr
from \citet{Garcia-Sanchez01}, one estimates that encounters by
stellar systems within 0.25 pc of the Sun occur with a frequency of
0.11 Myr$^{-1}$ or once every $\sim$9.2 Myr.  For comparison, flybys
this close (0.25 pc) are statistically rare ($\sim$2.4\%) among
encounters by all stellar systems that penetrate the Sun's tidal
radius of $\sim$1.35 pc \citep{Mamajek13}, of which $\sim$4.5 occur
per Myr \citep{Garcia-Sanchez01}.\\

\begin{figure}[htb!]
\plotone{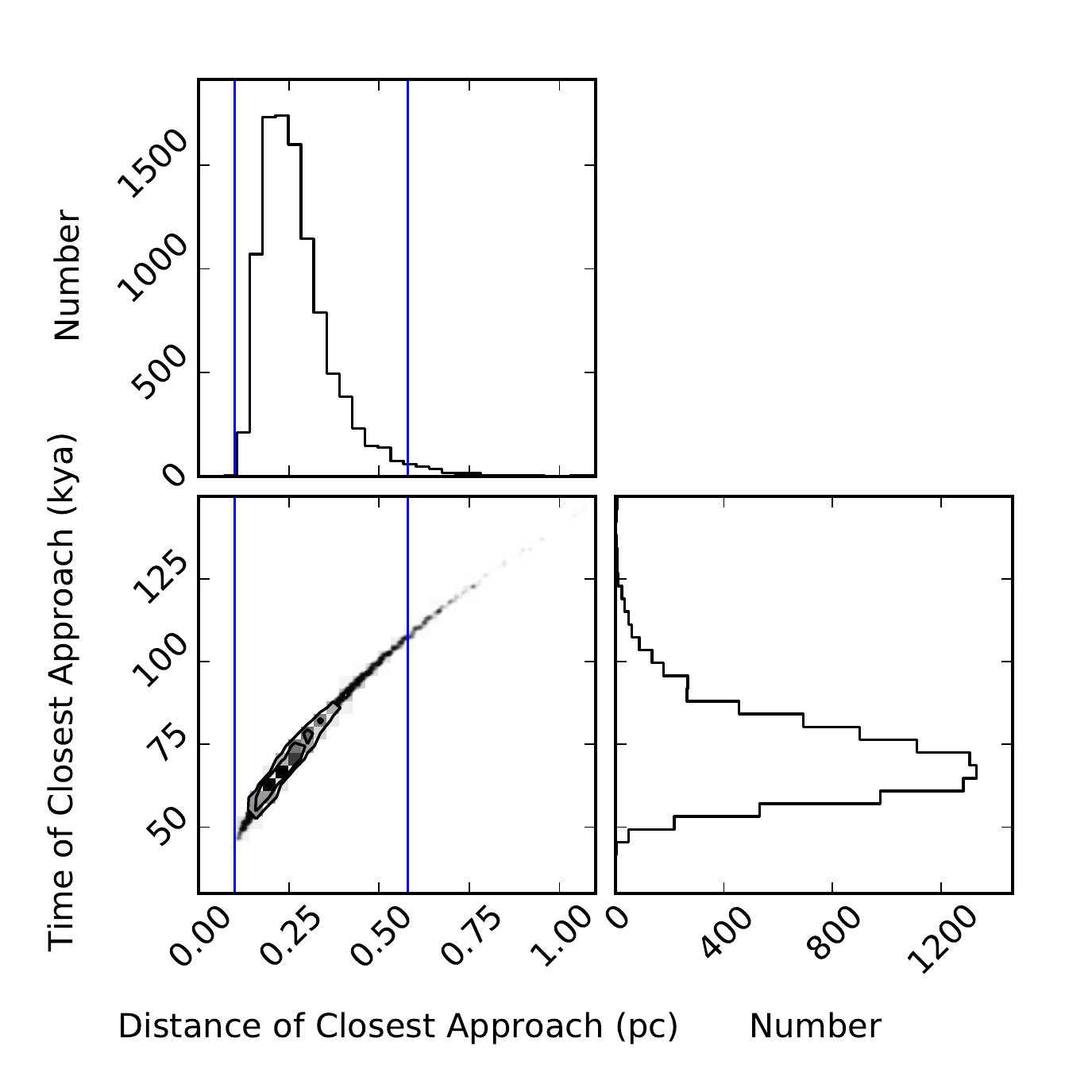}
\caption{{\it Lower left:} Density plot of the distribution of minimum
  separations between \name\, and the Sun (in pc) versus time of minimum
  separation (in thousands of years ago; ``kya'') for 10$^4$ Galactic
  orbit simulations. The maximum semi-major axis for retrograde
  orbiting Oort Cloud comets is $\sim$0.58 pc (right blue solid line),
  and the dynamically active inner Oort Cloud is at $<$0.10 pc (left
  blue solid line). The entire region plotted is within the Sun's
  tidal radius ($\sim$1.35 pc). {\it Upper left:} Histogram of
  simulated closest \name-Sun separations (pc).  {\it Lower right:}
  Histogram of simulated times of closest \name-Sun separation (kya).}
\end{figure}

The flyby's gravitational interaction is in the regime where the Sun's
influence on the star's trajectory (and vice versa) is negligible
\citep{Collins10}: $GM_{\odot}$/$bv^{2}_{*}$ $\simeq$ 10$^{-5.6}$ $<<$
1, where $G$ is the Newtonian constant, $M_{\odot}$ is the Sun's mass,
$b$ is impact parameter (minimum Sun-star separation), and $v_{*}$ is
the velocity of the star with respect to the solar system barycenter.
Major comet showers, where the flux of long-period comets increases by
factors of $>$10 are likely limited to cases of high mass interlopers
passing within $\sim$10 kAU of the Sun \citep{Heisler86}, and are
exceedingly rare \citep[$<$10$^{-3}$ Myr$^{-1}$; extrapolating from
  results by ][]{Garcia-Sanchez01}.  A perturbed comet with new
aphelion similar to that of the Sun-\name\, minimum separation, but
with perihelion in the planetary region of the solar system, will have
$a$ $\simeq$ 26 kAU and $P$ $\sim$ 4.2 Myr.  Although comets
throughout the Oort cloud may be perturbed by an interloper star onto
trajectories that would bring them to the inner solar system
\citep{Weissman96}, the highest density of outer Oort cloud comets
should exist at smaller orbital distances \citep{Heisler90}, so most
of the perturbed comets should originate from the vicinity of the
star's closest pass to the Sun.  Hence, any enhancement on the
long-period comet flux should become manifest $\sim$2 Myr in the
future.  A proxy indicator of the encounter-induced flux of Oort Cloud
comets is defined as $\gamma$ = $M_{*}$/$v_{*}b$ \citep[][using the
  variables as defined earlier, and with $M_{*}$ being the mass of the
  interloper binary \name]{Feng14}.  Using the 10$^4$ simulations, we
estimate that the \name\, flyby induced $\gamma$ $\sim$
10$^{-7.48\,\pm\,0.15 (1\sigma)}$
M$_{\odot}$\,km$^{-1}$\,s\,AU$^{-1}$.  Simulations by \citet{Feng14}
suggest that encounters with $\gamma$ $<$ 10$^{-5.3}$ are unlikely to
generate an enhancement in the distribution of longitudes for
long-period comets compared to that predicted to be generated by
Galactic tidal effects.  All of the 10$^4$ simulated orbits had
$\gamma$ $<$ 10$^{-7.0}$, hence the pass of the \name\, system should
have a negligible statistical impact on the flux of long-period comets
during the coming millenia.\\

\section{The Closest Known Flyby?}

The proximity of \name's flyby to the solar system can be compared to
those of \hip\, stars recently studied by \citet{Bailer-Jones14}.
He identified one star (HIP 85605) with a flyby closer than the median
minimum separation that we estimated for \name\, (0.25 pc), however
two future flybys (HIP 89825 [Gl 710] and HIP 63721) both have
$\Delta$ $\simeq$ 0.27 pc (0.27$^{+0.17}_{-0.17}$ pc and
0.27$^{+0.40}_{-0.23}$ pc, respectively; 90\%CL).
\citet{Bailer-Jones14} predicts that HIP 85605 will pass within
$\Delta$ = 0.10$^{+0.10}_{-0.06}$ pc (90\% CL) of the Sun $\sim$332
kyr in the future.  Unfortunately, as \citet{Bailer-Jones14} points
out, HIP 85605 is a visual binary, and there is considerable
dispersion in its published proper motions and its \hip\, parallax
is of low accuracy. The solution which brings HIP 85605 close to the
Sun relies on the Tycho-2 proper motion \citep[\pmra, \pmdec\, = +4.0,
  -7.6 $\pm$ 2.0, 1.9 mas\,yr$^{-1}$;][]{Hog00}, the revised \hip\,
parallax of $\varpi$ = 146.84\,$\pm$\,29.81 mas, and the Pulkovo
radial velocity of -21.0\,$\pm$\,0.3 km/s \citep[][as reported in the
  XHIP compiled catalog;
  \protect{\citet{Anderson12}}]{Gontcharov06}. If any of these values
are substantially in error, then the proximity of the flyby solution
is likely to be spurious. Both the original and revised \hip\,
astrometric solutions \citep{Perryman97, vanLeeuwen07} had much larger
proper motions than the Tycho-2 solution. \citet{Bailer-Jones14}
mentions that a reanalysis of the \hip\, astrometric solution by
F. van Leeuwen suggests ``{\it that the \hip-2 parallax of Hip
  85605 and its (relatively large) uncertainty are valid, but that on
  account of the large residuals and the complex nature of this system
  the solution should be treated with caution}.'' 

Unfortunately, the \hip\, parallax for HIP 85605 leads to some
astrophysical inconsistency.  As \citet{Bailer-Jones14} note, HIP
85605's B-V color (1.1 mag) is consistent with a K-type star, whereas
the absolute magnitude calculated using its revised \hip\, parallax
(M$_V$ = 11.86\,$\pm$\,0.45 mag) is more consistent with an M
dwarf. We find that HIP 85605's colors from \hip\, and 2MASS (B-V =
1.10\,$\pm$\,0.11, V-K$_s$ $\simeq$ 2.66\,$\pm$\,0.07, J-H $\simeq$
0.53\,$\pm$\,0.03, H-K$_s$ $\simeq$ 0.12\,$\pm$\,0.03) are all
consistent with a $\sim$K4 dwarf \citep[][]{Pecaut13}, in agreement
with the spectral template fitting analysis of \citet[][; which also
  yielded K4V]{Pickles10}. Dave Latham (priv. comm.) has visually
confirmed that five spectra of HIP 85605 taken with the CfA Digital
Speedometer \citep[as reported in][]{Garcia-Sanchez99} are indeed
consistent with a typical K dwarf.  As HIP 85605 is
spectrophotometrically a K dwarf, the \hip-derived absolute magnitude
(M$_V$ $\simeq$ 11.9) {\it places the star nearly 5 magnitudes fainter
  than the main sequence} (M$_V^{MS}$ $\simeq$ 7.0) for a star of its
color - too faint to be simply a metal-poor dwarf, yet too red and
luminous for a white dwarf\footnote{See RECONS HR diagram for 10 pc
  sample (http://www.recons.org/hrd.2010.0.html) and
  http://dx.doi.org/10.6084/m9.figshare.1284334}. Hence, most likely
HIP 85605 is a $\sim$K4V at $d$ $\simeq$ 60 pc and the \hip\, parallax
is erroneous.  At $d$ $\simeq$ 60 pc, HIP 85605 has a slightly
different velocity ($U, V, W$ = -10.7, -14.3, -11.6 km\,s$^{-1}$)
compared to that used in the simulation by \citet{Bailer-Jones14}, and
its ``flyby'' of the Sun correspondingly moves further into the future
and further away ($\sim$2.8 Myr in future, $\sim$10 pc away). We
conclude that HIP 85605 is unlikely to penetrate the Sun's Oort Cloud,
and that \name\, now appears to have the closest flyby of any known
star.

\section{Discussion\label{sec:discussion}}

Given its current visual magnitude of $V$ $\simeq$ 18.3
\citep{Ivanov14}, at its closest approach of $\sim$0.25 pc, Scholz's
star (\name) would have had an apparent magnitude of $V$ $\simeq$
10.3, brighter than the current nearest star (Proxima; $V$ = 11.2) but
still much dimmer than the faintest naked eye stars ($V$ $\sim$
6). However, \name\, is an active M dwarf star \citep{Burgasser14},
and $V$-band flares have been observed to exceed 9 mag on timescales
of minutes amongst such stars, and brief flares exceeding 12 mag may
be possible \citep{Schmidt14}. Flares amongst the coolest M dwarfs
have been witnessed with energies of $\sim$10$^{34}$ ergs
\citep{Schmidt14} and luminosities of $\sim$10$^{29}$-10$^{30}$
erg\,s$^{-1}$. If \name\, experienced occasional flares similar to
those of the active M8 star SDSS\,J022116.84+194020.4
\citep{Schmidt14}, then the star may have been rarely visible with the
naked eye from Earth ($V < 6$; $\Delta V < -4$) for minutes or hours
during the flare events.  Hence, while the binary system was too dim
to see with the naked eye in its quiescent state during its flyby of
the solar system $\sim$70 kya, flares by the M9.5 primary may have
provided visible short-lived transients visble to our ancestors.

Improved astrometry for \name\, via ground-based telescopes
\citep{Dieterich14} or \emph{Gaia} \citep{deBruijne12}, and further
radial velocity monitoring, should help reduce the uncertainties in
the flyby timing and minimum separation between Scholz's star and the
solar system. Past systematic searches for stars with close flybys to
the solar system have been understandably focused on the \hip\,
astrometric catalog \citep{Garcia-Sanchez99, Bailer-Jones14}, however
it contains relatively few M dwarfs relative to their cosmic
abundance. Searches in the \emph{Gaia} astrometric catalog for nearby
M dwarfs with small proper motions and large parallaxes (i.e. with
small tangential velocities) will likely yield addition candidates.

\acknowledgements

EEM acknowledges support from NSF grant AST-1313029. AYK and PV
acknowledge the support from the National Research Foundation (NRF) of
South Africa. We thank Dave Latham, Alice Quillen, Kevin Luhman,
Cameron Bell, Dave Cameron, Peter Teuben, Segev Ben-Zvi, Adam
Burgasser, Ralf-Dieter Scholz, Matt Multunas, and Richard Sarkis for
discussions. We thank the referee for a very timely and useful report.

\bibliographystyle{apj}

\end{document}